\documentclass[%
 reprint,
 amsmath,amssymb,
 aps,
]{revtex4-2}

\usepackage{graphicx}% Include figure files
\usepackage{dcolumn}% Align table columns on decimal point
\usepackage{bm}% bold math
\usepackage{amsmath,amsthm,amssymb}
\usepackage{enumitem}
\def\be{\begin{equation}}
\def\ee{\end{equation}}

\DeclareMathOperator{\sech}{sech}

\begin{document}

\title{
Radiation damping of the soliton internal mode\\ in 1D quadratic Klein-Gordon equation}
%\thanks{A footnote to the article title}%

\author{Piotr Bizo\'n}
\author{Tomasz Roma\'nczukiewicz}
\affiliation{Institute of Theoretical Physics, Jagiellonian University, Krak\'ow}
\email{piotr.bizon@uj.edu.pl, tomasz.romanczukiewicz@uj.edu.pl}

\date{\today}% It is always \today, today,
             %  but any date may be explicitly specified

\begin{abstract}
We study long-time dynamics of small even perturbations of the soliton in 1D quadratic nonlinear Klein–Gordon equation. The soliton  possesses both an internal mode and an unstable mode. On a codimension-one manifold of fine-tuned initial data, the instability is suppressed and the internal mode decays slowly by transferring energy into the continuum. We show that this decay and the associated nonlinear frequency shift are accurately captured by a cubic resonant  approximation, with the damping rate determined by a Fermi golden rule-type coefficient. This provides a quantitative description of irreversible energy transfer from the internal mode to dispersive radiation.
\end{abstract}

%\keywords{Suggested keywords}%Use showkeys class option if keyword
                              %display desired
\maketitle

{\bf Introduction.}
Many nonlinear wave equations supporting solitary waves feature internal modes, i.e. localized oscillatory excitations of the soliton. These modes play a crucial role in the soliton’s long-time dynamics, mediating energy exchange between the coherent core and the  continuum. Generically, internal modes are long-lived, but nonlinear interactions with the continuum induce slow radiation damping, leading to gradual decay of the mode amplitude and a small frequency shift~\cite{SW}. Understanding this mechanism of metastability is important in contexts ranging from optical fibers and Bose–Einstein condensates to field-theoretic solitons.

In this paper we analyze the radiation damping of the soliton internal mode for  the quadratic Klein-Gordon equation in one spatial dimension,
\be\label{kg2}
\phi_{tt}-\phi_{xx} +\phi = \phi^2.
\ee
This equation possesses a static finite-energy solution, which we refer to as the soliton
\be\label{S}
\phi=S(x)=\frac{3}{2} \sech^2\left(\frac{x}{2}\right).
\ee
Substituting  $\phi(t,x)=S(x)+u(t,x)$ into Eq.~\eqref{kg2} we obtain the equation for the perturbation $u(t,x)$
\be \label{equ}
u_{tt} +L u = u^2, \qquad L=-\partial_{x}^2-3 \sech^2\left(\frac{x}{2}\right)+1\,.
\ee
The  operator $L$  has continuous spectrum $[1,\infty)$ and the following orthonormal eigenfunctions and eigenvalues:
\begin{description}[font=\normalfont]
\item[Unstable mode] $L\xi=-\lambda^2 \xi$,
\begin{equation}\label{xi}
\xi(x)=\sqrt{\frac{15}{32}}\,\sech^3\left(\frac{x}{2}\right), \quad \lambda^2=\frac{5}{4}\,.
\end{equation}

\item[Zero mode] $L\chi=0$,
\begin{equation}\label{chi}
\chi(x)=\sqrt{\frac{15}{8}}\,\tanh\left(\frac{x}{2}\right)\sech^2\left(\frac{x}{2}\right)\,.
\end{equation}

\item[Internal mode] $L\psi=\omega^2 \psi$,
\begin{equation}\label{psi}
\psi(x)=\sqrt{\frac{3}{32}}\,\frac{1-4\sinh^2\left(\frac{x}{2}\right)}{\cosh^3\left(\frac{x}{2}\right)}, \quad \omega^2=\frac{3}{4}\,.
\end{equation}
\end{description}
\vspace{-0.25cm}
The coexistence of an internal mode with an unstable  mode makes the one-dimensional quadratic Klein–Gordon equation a paradigmatic setting for studying radiation-induced relaxation in the presence of an unstable direction.
For simplicity, we restrict our analysis to even initial data, so the odd zero mode (associated with invariance under spatial translations) will not be excited. This also avoids technical complications arising from the odd resonance at the edge of the continuous spectrum.

It is well known that for small initial data solutions of Eq.~\eqref{kg2}  exist globally in time, while sufficiently large initial data can result in finite-time singularities. It was shown numerically in \cite{bcs} that even solutions fine-tuned to the  threshold of blowup approach 
the soliton through a slow decay of the internal mode amplitude,
however the rate of convergence was not determined (see also \cite{MR} for the collective coordinates approach). 
Recently, Li and L\"uhrmann \cite{LL} established the codimension-one asymptotic stability of $S$ for small even perturbations, 
using virial-type arguments pioneered by Kowalczyk, Martel, and Mu\~noz (see \cite{KMM1} and the survey \cite{KMM2}). 
This method provides an integrated-in-time decay estimate but does not yield an explicit rate. 

The mechanism of decay, originally proposed  by Sigal \cite{S}, was rigorously analyzed by Soffer and Weinstein in the seminal work \cite{SW}  for the cubic  Klein–Gordon equation with a potential. They showed that the nonlinear resonant interaction between the discrete bound state and the continuous spectrum leads to a transfer of energy from the bound state to dispersive radiation.

 In our case, the dynamics of the internal mode is
interwined  both with the radiation and the unstable mode; consequently, relaxation to the soliton
can occur only if the unstable direction is suppressed. Under this assumption, we employ
normal form methods to derive a cubic resonant approximation for the dynamics of the internal mode. This provides an explicit decay rate for the internal mode amplitude,
together with the associated nonlinear frequency shift. These analytical results are
corroborated by numerical computations.
%The agreement between formal analytic and numerical computations makes us  feel confident that the results are true.
%
\vskip 1cm
{\bf Cubic resonant approximation.}
We begin by decomposing even perturbations of $S$ into mutually orthogonal spectral components
\be\label{dec}
u(t,x)=a(t) \psi(x) + b(t) \xi(x) +\eta(t,x)
\ee
with  $\langle \psi,\xi \rangle=\langle \psi,\eta \rangle= \langle \xi,\eta \rangle=0$, where $\langle f,g \rangle$
% := \int_{-\infty}^{\infty} f(x) g(x) dx $
 denotes the standard inner  product on~$L^2$.
Substituting this decomposition into Eq.~\eqref{equ}, we obtain a coupled system of 
two ordinary differential equations (ODEs) for the coefficients $a(t)$ and $b(t)$ of the internal and unstable modes, 
and the PDE for the radiation field $\eta(t,x)$,
\begin{eqnarray} 
\ddot a +\omega^2 a & = & \langle u^2,\psi \rangle, \label{eqa}\\
\ddot b -\lambda^2 b & = &  \langle u^2,\xi \rangle,\label{eqb} \\ 
\eta_{tt} +L \eta &=& P_c(u^2) \label{eqeta}\,.
\end{eqnarray}
Here $P_c(f) := f-\langle \xi, f \rangle \xi- \langle \psi, f \rangle \psi$ denotes the projection of function $f$  onto the continuous spectrum of $L$.
\vskip 0.2cm
Let us temporarily neglect the radiation field $\eta$. The system \eqref{eqa}-\eqref{eqeta} then reduces to two ODEs
\begin{eqnarray}
\ddot a +\omega^2 a & = & c_1 a^2 + 2 c_2 a b + c_3 b^2,\label{odea}\\
\ddot b -\lambda^2 b & = & c_2 a^2 + 2 c_3 a b + c_4 b^2,\label{odeb}
\end{eqnarray}
with the  coefficients $c_j$  given by
\begin{align*}
&c_1=\int \psi^3 = \sqrt{\frac{3}{2}} \frac{603\pi}{8192},  & c_2= \int \psi^2 \xi=\sqrt{\frac{15}{2}} \frac{129\pi}{8192}, \\& c_3=\int \psi \xi^2= -\sqrt{\frac{3}{2}} \frac{225\pi}{8192},  &c_4=\int \xi^3=\sqrt{\frac{15}{2}} \frac{525\pi}{8192}\,.
\end{align*}
This reduced system is hamiltonian and can be alternatively derived using collective coordinates i.e. evaluating the Lagrangian functional for Eq.~\eqref{equ} on the ansatz \eqref{dec} with $\eta$ set to zero and integrating over $x$; see \cite{MR}.

The origin $a=b=0$ is a saddle-center fixed point. 
By the Lyapunov center theorem, a continuum of periodic solutions bifurcates from it. 
Assuming that $a=\mathcal{O}(\varepsilon)$ and $b=\mathcal{O}(\varepsilon^2)$ for a small $\varepsilon$, these periodic solutions and their stable/unstable manifolds admit convergent perturbative expansions which can be constructed by iterating normal form transformations \cite{M,G}.
For our purposes, it suffices to  eliminate quadratic terms from the right-hand sides of Eqs.~\eqref{eqa} and \eqref{eqb}. To simplify notation below $\mathcal{O}(\varepsilon^k)$  will be abbreviated as $\mathcal{O}_k$. 

We  define the complex variable
$
z = a + \frac{i}{\omega} \dot a
$,
so that
$
a = \frac{1}{2} (z + \bar z)$.
In terms of $z$, Eq.~\eqref{odea} becomes
\begin{equation} \label{odez}
\dot z + i\omega z = \frac{i}{\omega} \left(c_1 a^2 + 2 c_2 a b + c_3 b^2\right).
\end{equation}
First, we make  a near-identity transformation
\begin{equation} \label{normal_b}
b = \tilde b + \beta_1 (z^2 + \bar z^2) + \beta_2 |z|^2.
\end{equation}
Substituting \eqref{normal_b} into Eq.~\eqref{odeb}, using \eqref{odez}, and choosing
\begin{equation}
\beta_1 = -\frac{c_2}{16\omega^2 + 4\lambda^2} = -\frac{c_2}{17}, \quad 
\beta_2 = -\frac{c_2}{2\lambda^2} = - \frac{2 c_2}{5}, \nonumber
\end{equation}
we eliminate the $\mathcal{O}_2$ term $c_2 a^2$ on the right-hand side. It follows that 
$\tilde b(t) = \mathcal{O}_3$, and setting $z(0) = \varepsilon$ we obtain on the stable manifold
\begin{equation} \label{b_approx}
b(t) = C e^{-\lambda t} - \frac{2 c_2 \varepsilon^2}{85} \left(17 + 5 \cos(2\omega t)\right)  + \mathcal{O}_3,
\end{equation}
where $C$ is a free parameter. The exponentially decaying term is negligible for late times and will be omitted below.

Next, we turn to Eq.~\eqref{odez}. It is easy to see that the near-identity transformation
\begin{equation} \label{normal_z}
z = y + \frac{c_1}{12\omega^2} (-3 y^2+ \bar{y}^2 + 6 |y|^2)
\end{equation}
removes the quadratic term and produces the first cubic resonant contribution
\begin{equation} \label{reson2}
i \frac{c_1}{\omega} a^2 \longrightarrow i \frac{5 c_1^2}{12 \omega^3}\, y |y|^2, 
\end{equation}
up to cubic nonresonant terms and $\mathcal{O}_4$.
The second cubic resonant contribution arises from 
 substituting \eqref{normal_b} into the second term on the right-hand side of \eqref{odez}, yielding
\begin{equation} \label{reson1}
i \frac{2 c_2}{\omega} a b \longrightarrow - i \frac{39 c_2^2}{85 \omega}\, y |y|^2,
\end{equation}
again up to cubic nonresonant terms and $\mathcal{O}_4$.
Combining \eqref{reson2} and \eqref{reson1}, and dropping nonresonant and higher-order terms, we obtain
\begin{equation} \label{ode_y}
\dot y + i \omega y = i \gamma_1 y |y|^2,
\end{equation}
where
\begin{equation} \label{gamma1}
\gamma_1 = \frac{5 c_1^2}{12 \omega^3} - \frac{39 c_2^2}{85 \omega} \approx 0.041732.
\end{equation}
Finally, we set
$y = A(t) e^{-i\omega t}
$
to filter out the fast linear oscillations. This yields the cubic Birkhoff normal form associated with the internal mode
\begin{equation} \label{reA}
\dot A = i \gamma_1 A |A|^2.
\end{equation}
Since the coefficient $i \gamma_1$ is purely imaginary, the cubic resonant term induces only a nonlinear frequency shift and no decay, reflecting the hamiltonian character of the finite-dimensional truncation.
For $A(0) = \varepsilon$, this gives
\[
A(t) = \varepsilon e^{i \varepsilon^2 \gamma_1 t},
\]
which, using \eqref{normal_z}, translates into the following second-order approximation for  the internal mode
\begin{equation} \label{a_approx}
a(t) = \varepsilon \cos(\tilde \omega t) + \frac{c_1 \varepsilon^2}{6 \omega^2} \left(3 - \cos(2 \tilde \omega t)\right) + \mathcal{O}_3,
\end{equation}
where $\tilde \omega = \omega - \gamma_1 \varepsilon^2 + \mathcal{O}_4$ is the renormalized frequency.
\vskip 0.1cm
We now reinstate the radiation field $\eta$ and analyze the full infinite-dimensional Hamiltonian system \eqref{eqa}--\eqref{eqeta}. The PDE~\eqref{eqeta} can be written as 
\begin{equation} \label{eqeta2}
\eta_{tt} + L \eta = \frac{1}{4} (z + \bar z)^2 P_c(\psi^2) + \mathcal{O}_3.
\end{equation}
To remove the quadratic term from the right-hand side, we introduce a near-identity transformation
\begin{equation} \label{normal_eta}
\eta = \tilde \eta + f_1(x) z^2 + \bar f_1(x) \bar z^2 + f_2(x) |z|^2.
\end{equation}
Substituting \eqref{normal_eta} into Eq.~\eqref{eqeta2} yields
\begin{align} \label{etatilde}
\tilde \eta_{tt} &+ L \tilde \eta
+ (L - 4\omega^2)\bigl( z^2 f_1 + \bar z^2 \bar f_1 \bigr)
+ |z|^2 L f_2 \nonumber \\
&= \frac{1}{4} (z + \bar z)^2 P_c(\psi^2) + \mathcal{O}_3.
\end{align}
This leads to the following equations for $f_1$ and $f_2$
\begin{align}
(L - 4\omega^2) f_1 &= \frac{1}{4} P_c(\psi^2), \label{eqf1} \\
L f_2 &= \frac{1}{2} P_c(\psi^2), \label{eqf2}
\end{align}
which ensure that the quadratic terms in \eqref{etatilde} vanish, and therefore $\tilde \eta = \mathcal{O}_3$.

Eq.~\eqref{eqf2} has a unique real-valued solution that vanishes as $|x| \to \infty$. 
%The explicit expression is given in the Supplemental Material (SM).
This solution can be found by the method of reduction of order and the substitution $f_2(x) = \chi(x) h(x)$, 
where $\chi(x)$ is the zero mode. The explicit formula is somewhat unwieldy, so we do not record it here.
\vskip 0.1cm

Since $4\omega^2 = 3 > 1$, the frequency $2\omega$ generated by the quadratic nonlinearity lies within the continuous spectrum. 
Consequently, solutions of Eq.~\eqref{eqf1} oscillate at infinity. The  physically relevant solution is singled out by imposing the outgoing boundary conditions as $|x| \to \infty$. Using the method of variation of parameters, we write this
unique complex-valued solution in the form
\begin{align} \label{c1}
f_1(x)
&= \frac{1}{4 W(p)} \Biggl[
   j_{+}(x,p) \int_{-\infty}^x j_{-}(s,p) P_c(\psi^2(s))\, ds \notag \\
&\qquad + j_{-}(x,p) \int_{x}^{\infty} j_{+}(s,p) P_c(\psi^2(s))\, ds
\Biggr]\,,
\end{align}
where $p = \sqrt{4\omega^2 - 1} = \sqrt{2}$ and $j_{\pm}(x,k)$ are the Jost solutions of the Schrödinger equation  
\[
(L-1) j_{\pm} = k^2 j_{\pm},
\]  
satisfying $j_{\pm}(x,k) \sim e^{\pm i k x}$ as $x \to \pm\infty$. The Wronskian of $j_+$ and $j_-$ is denoted by $W(k)$; see the Supplemental Material for  details.

In the derivation of the cubic resonant approximation~\eqref{reA}, we temporarily
neglected the radiation field~$\eta$.
Since both $b$ and $\eta$ are of order $\mathcal{O}_2$, the leading cubic term
omitted from the right-hand side of Eq.~\eqref{eqa} is
$
2 a \langle \psi \eta, \psi \rangle.
$
Restoring this term and substituting the near-identity transformation~\eqref{normal_eta} for $\eta$, we obtain
 two additional cubic resonant terms on the right-hand side of
Eq.~\eqref{reA},
\begin{equation} \label{reA_eta}
\frac{i}{\omega}
\Bigl( \langle \psi f_1, \psi \rangle
      + \langle \psi f_2, \psi \rangle \Bigr)
A |A|^2.
\end{equation}
The coefficient $\frac{i}{\omega} \langle \psi f_2, \psi \rangle$ is purely imaginary and therefore only modifies the phase. 
However -- and this is the key point -- the coefficient $\frac{i}{\omega} \langle \psi f_1, \psi \rangle$ has a negative real part, which induces damping.

Using Maple we find
\begin{align*}
\frac{1}{\omega} \langle \psi f_2, \psi \rangle &:= \gamma_2 
= \frac{2}{\sqrt{3}} \left( \frac{11}{224} - \frac{156843 \pi^2}{33554432} \right) 
\approx 0.003433, \\[2mm]
\frac{i}{\omega} \langle \psi f_1, \psi \rangle &:= -\frac{\Gamma}{2} + i \gamma_3,
\end{align*}
where $\gamma_3\approx 0.000772$ and
\be
\Gamma = \frac{1}{2 p \omega} \left| \langle j_+(\cdot, p), \psi^2 \rangle \right|^2
= \frac{729 \sqrt{6}\, \pi^2}{1088 \sinh^2(\sqrt{2}\pi)} \approx 0.008966.
\ee
The condition $\Gamma > 0$, known as the Fermi golden rule, ensures a nonvanishing resonant coupling between the internal mode and the continuum at the cubic order; see \cite{LL} for an elegant computation of $\Gamma$ using  supersymmetric factorization properties of the operator $L$.

Putting all the above pieces  together, we obtain the cubic resonant approximation for the dynamics of the internal mode
\begin{equation} \label{reA2}
\dot A = \bigl( i \gamma - \Gamma/2 \bigr) A |A|^2,
\end{equation}
where 
$\gamma = \gamma_1 + \gamma_2 + \gamma_3 \approx 0.045938.
$
Unlike the hamiltonian normal form \eqref{reA}, this equation is \emph{dissipative}: the negative real part of the cubic coefficient accounts for irreversible energy transfer from the internal mode to dispersive radiation and results in a slow decay of the amplitude.
Plugging the polar representation $A(t)=R(t)e^{i\theta(t)}$ into \eqref{reA2}, we obtain the amplitude and phase equations
\begin{equation} \label{amplitude-phase}
\dot R = -(\Gamma/2) R^3, 
\qquad 
\dot{\theta} = \gamma R^2.
\end{equation}
Solving these equations with initial condition $A(0)=\varepsilon$, we find
\begin{equation} \label{R}
R(t) = \frac{\varepsilon}{\sqrt{1 + \varepsilon^2 \Gamma t}}, 
\qquad
\theta(t) = \frac{\gamma}{\Gamma} \ln \bigl( 1 + \varepsilon^2 \Gamma t \bigr).
\end{equation}
Notice that for  $t\gg \varepsilon^{-2}$, the internal mode amplitude $R(t) \approx \frac{1}{\sqrt{\Gamma}} \ t^{-1/2}$  and the   phase shift $\theta(t) \approx (\gamma/\Gamma) \ln{t}$, are independent of $\varepsilon$. Thus, the asymptotic behavior of the internal mode is universal: the smallness of the initial amplitude is forgotten and the profile is determined solely by the normal form coefficients.
\vskip 0.1cm
{\bf Numerical validation.} To verify the prediction \eqref{R}, we perform direct numerical simulations of Eq.~\eqref{equ} with initial data
\be \label{in_data}
u(0,x)=\varepsilon \psi(x)+b_0 \xi(x), \qquad u_t(0,x)=0,
\ee
where the parameter $b_0$ is fine-tuned via shooting to the blowup/dispersion threshold (see \eqref{b_approx} for the second-order approximation of $b_0$). The timescale $T$ over which the solution remains close to the codimension-one stable manifold is limited by the exponential growth of numerical errors along the unstable direction. For double-precision arithmetic this gives $T \sim \frac{1}{\lambda} \ln(10^{16})~\approx 33$. 
To overcome this limitation, we employ a repeated shooting-to-threshold procedure: after evolving over intervals of $T_{\rm corr} \approx 20$, a small correction $\delta b$ along the unstable eigenfunction is determined via shooting to restore the solution to the blowup/dispersion threshold. The value of $T_{\rm corr}$ was chosen empirically to ensure numerical stability while minimizing the number of corrections. This procedure allows us to track the evolution along the codimension-one stable manifold up to times $ \sim 5000$.

To compare the numerical evolution with the formula \eqref{R}, we monitor the field at a fixed spatial point $x_0$ and record its values $u(t_j,x_0)$ at the times $t_j$ corresponding to successive local maxima. 
We then fit the sequence $\left(\frac{\psi(x_0)}{u(t_j,x_0)}\right)^2$ to the theoretical prediction
$
\varepsilon^{-2} + \Gamma t.
$
To verify the  frequency shift, we compute
the instantaneous frequency  via
$
\omega_{\mathrm{inst}}(t_j) \approx \frac{2\pi}{t_{j+1}-t_j}.
$
Since $\omega_{\mathrm{inst}}(t)=\omega-\dot\theta(t)$, this provides a numerical approximation of $\dot\theta(t)$. We then fit the sequence $1/\dot{\theta}(t_j)$
to the theoretical prediction $\frac{1}{\gamma\varepsilon^2} + \frac{\Gamma}{\gamma} t$.
For a sample value $\varepsilon=0.2$, these fits over the interval $500<t<5000$ give  $\Gamma \approx 0.009011$ and
    $\gamma\approx 0.04564$ in agreement with the  theory to within  $1\%$.
    
    \begin{figure}[t]
    \centering
   \includegraphics[width=0.8\columnwidth]{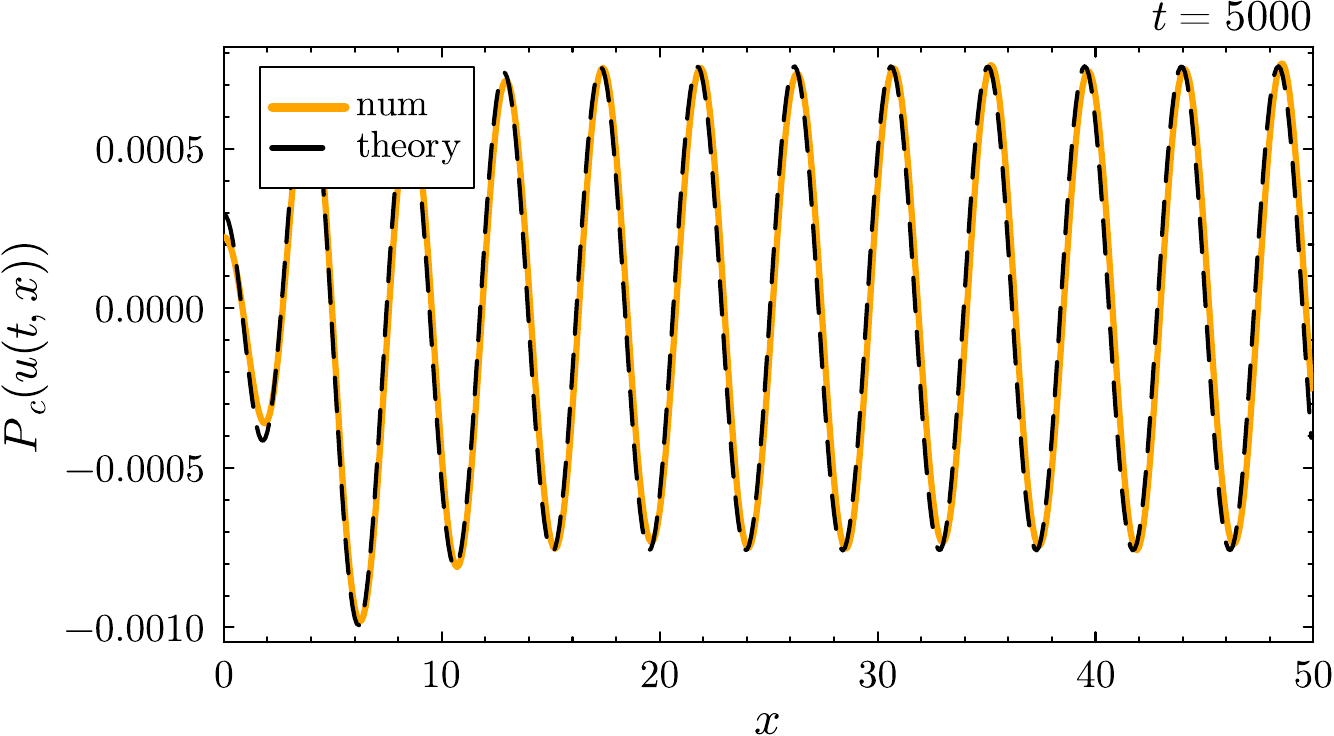}\\
   \vspace{0.2cm}
    \includegraphics[width=0.8\columnwidth]{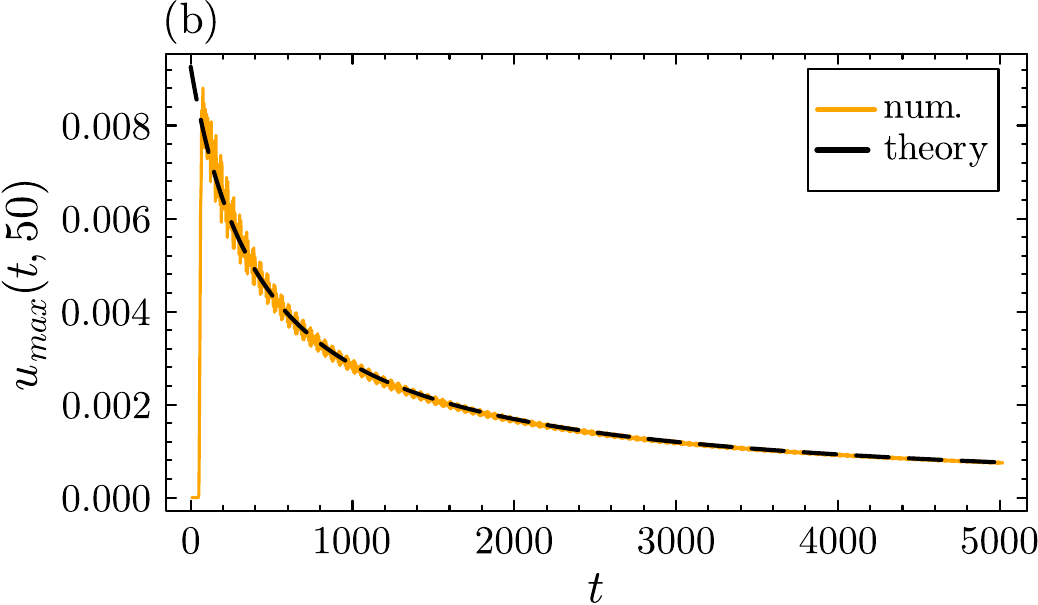}
    \caption{Numerical evolution of  initial data \eqref{in_data} with $\varepsilon=0.5$. (a) Spatial profile $\eta(5000,x)$. (b) Time profile $\eta(t,50)$. The approximation \eqref{eta_approx} is superimposed for comparison. In (a) the theoretical result is plotted for a slightly shifted time, 5000.76, to offset a small phase drift.}
    \label{fig:eta}
\end{figure}
\vskip 0.1cm
{\bf Internal mode feedback on radiation.}
Substituting $z=R e^{i(\theta-\omega t)}$, with $R(t)$ and $\theta(t)$ given in \eqref{R}, into \eqref{normal_eta},  we obtain a rough second-order approximation of the  internal mode backreaction  on  radiation
 \be \label{eta_approx}
\eta_{\mathrm{app}}(t,x) = R^2(t)\bigl[f_1(x) e^{-2i(\omega t-\theta(t))}  
+\text{c.c.} + f_2(x)\bigr].
\ee
The term with $f_1(x)$
 represents  two outgoing distorted plane waves with wavenumbers $k=\pm p=\pm \sqrt{2}$ and frequency $2\omega=\sqrt{3}$ (hence phase velocity $v_{\textrm{ph}}=\pm\sqrt{3/2}$) emitted by the internal mode oscillations, while the term with $f_2(x)$ is a localized non-radiative distortion.
 For large $|x|$ and $t$, \eqref{eta_approx} reduces to
\begin{equation} \label{eta_decay}
\eta_{\mathrm{app}}(t,x) \approx c\, t^{-1}
\cos\bigl(\sqrt{3} t \pm \sqrt{2} x - \frac{2\gamma}{\Gamma}\log t\bigr),
\end{equation}
where
$
c=\frac{6^{\frac{1}{4}}}{4 \sqrt{\Gamma}}.
$
Here we used $f_1(x)\approx \frac{6^{\frac{1}{4}}\sqrt{\Gamma}}{8} e^{\pm i \sqrt{2} x}$ and set $f_2(x)$ to zero.
Note that this decay  is slower than the $t^{-3/2}$ decay of linear dispersive waves for even initial data. 

Although \eqref{eta_approx} compares reasonably well with numerical evolution (see Fig.~\ref{fig:eta}), we stress that, in order to properly compute the leading-order backreaction, one should solve the radiation equation
\begin{equation} \label{eqeta3}
\eta_{tt} + L \eta = \frac{1}{4} (z + \bar z)^2 P_c(\psi^2),
\end{equation}
with $z = R(t) e^{i(\theta(t) - \omega t)}$ and $R(t)$ and $\theta(t)$ given in \eqref{R}. This  computation, to be reported elsewhere, predicts  a logarithmically modulated decay  $ t^{-1/2}$ for wave packets propagating with group velocity $v_{\textrm{g}}=\pm\sqrt{2/3}$. The slow-down effect in comparison with \eqref{eta_decay} arises from the near  time-resonance in the distorted Fourier transform corresponding to the frequency $\sqrt{k^2+1}=2\omega=\sqrt{3}$.
The  logarithmic phase modulation detunes the exact time resonance, preventing  the   growth factor $\log{t}$ which would otherwise appear along the  resonant ray \cite{JL}. Similar phenomena have been analyzed  for the $\phi^4$ model \cite{DM}  and  the 3D quadratic Klein–Gordon equation with a potential \cite{LP1,LP2}; see also \cite{LLS,LLSS,LS,BDKK} for related work.
\vskip 0.1cm
{\bf Outlook.}
We view the 1D quadratic nonlinear Klein--Gordon equation as an infinite-dimensional Hamiltonian system whose phase space splits, near the soliton, into two discrete degrees of freedom (corresponding to the internal and unstable modes) coupled to a continuum of dispersive waves. On a codimension-one stable manifold, the instability is suppressed, and the internal mode undergoes slow, irreversible decay by radiating energy into the dispersive bath. Using normal form techniques, we derive an effective cubic resonant description that captures this damping and the associated nonlinear frequency shift, in excellent agreement with numerical simulations. 

In physical terms, the internal mode acts as a small antenna attached to the soliton, gradually leaking energy into dispersive waves that travel away from the core.  Since internal modes are ubiquitous in nonlinear wave systems, this mechanism of relaxation is expected to arise in many physical contexts.  Understanding the coupling between localized dynamics and dispersive radiation may therefore help clarify relaxation and energy-transfer mechanisms in a broad class of nonlinear media.

\vskip 0.1cm
{\it Acknowledgments.} We thank J.~Jendrej, M.~Kowalczyk, and J.~L\"uhrmann for helpful discussions. PB was supported  by the Polish National Science Center grant no. 2017/26/A/ST2/00530.

\newpage
\onecolumngrid
\section*{Supplemental Material}

\subsection{Jost functions}
We construct the Jost functions of the Schr\"odinger operator $L-1$ using ladder operators, for $\ell = 1,2,3$, 
\[
\mathcal{D}_\ell = \partial_x + \frac{\ell}{2} \tanh\frac{x}{2}.
\]
These operators satisfy the conjugation identity
\[
\mathcal{D}_1 \mathcal{D}_2 \mathcal{D}_3 (L-1) = -\partial_x^2 \, \mathcal{D}_1 \mathcal{D}_2 \mathcal{D}_3.
\]
Taking the adjoint and acting on plane waves, we obtain
\[
(L-1) \, \mathcal{D}_3^\star \mathcal{D}_2^\star \mathcal{D}_1^\star (e^{ikx}) = k^2 \, \mathcal{D}_3^\star \mathcal{D}_2^\star \mathcal{D}_1^\star (e^{ikx}).
\]
Thus the Jost functions are
\[
j_\pm(x,k) = \pm \nu(k) \, \mathcal{D}_3^\star \mathcal{D}_2^\star \mathcal{D}_1^\star \bigl(e^{\pm ikx}\bigr),
\]
where $\nu(k)$ is a normalization factor chosen so that
$
\lim_{x \to \pm \infty} j_\pm(x,k) \, e^{\mp ikx} = 1.
$
Using the shorthand $y = \tanh(x/2)$, the Jost functions can be written explicitly as
\begin{align}\label{jost_explicit}
j_{\pm}(x,k) &= \nu(k) \Bigl[ i k^3 
\mp 3 y k^2 
- i k \Bigl(\frac{15}{4} y^2 - 1\Bigr)  \pm \frac{15}{8} y \Bigl(y^2 - \frac{3}{5}\Bigr) \Bigr] e^{\pm i k x},
\end{align}
where the normalization factor is
\begin{align}
\nu(k) &= \Bigl(i k^3 - 3 k^2 - \frac{11}{4} i k + \frac{3}{4}\Bigr)^{-1}.
\end{align}
It follows from \eqref{jost_explicit} that the reflection coefficient vanishes and the transmission coefficient is
\begin{align}
T(k) &= \frac{4 i k^3 - 12 k^2 - 11 i k + 3}
           {4 i k^3 + 12 k^2 - 11 i k - 3},
\end{align}
while the Wronskian is
\begin{align}
W(k) &:= j_+ \partial_x j_- - j_- \partial_x j_+ 
      = -2 i k \, T^{-1}(k).
\end{align}

\end{document}